\documentclass[conference]{IEEEtran}%% Packages / general settings
\usepackage{graphicx}
\usepackage{amsmath,amssymb,dsfont,bbm,epstopdf,pgfplots,mathtools,enumitem,mathrsfs, bbm, algpseudocode , graphicx, dblfloatfix, booktabs, physics, algorithm,algcompatible, derivative} 
\usepackage[normalem]{ulem}
\usepackage{color}
\usepackage{cite}%hyperref}
\usepackage{graphics}
\usepackage{tikz}
\usepackage{pgfplots}
\usepackage{subcaption}
\usepackage[utf8]{inputenc} % usually not needed (loaded by default)
\usepackage[T1]{fontenc}
\usepackage[left=0.6in,right=0.6in,top=0.71in, bottom = 0.94in]{geometry}
\usetikzlibrary{shapes, arrows, decorations.markings, arrows.meta}
\usetikzlibrary{patterns} % LATEX and plain TEX when using Tik Z
\allowdisplaybreaks
\graphicspath{{.}{./Figures/}} % path to figures
\allowdisplaybreaks[4] % Allows the align environment to wrap over pages.

\newtheorem{theorem}{Theorem}
\newtheorem{proposition}{Proposition}
\newtheorem{remark}{Remark}
\newtheorem{definition}{Definition}

\renewcommand{\Pr}{{\mathbb{P}}}

\definecolor{ForestGreen}{rgb}{0.0, 0.5, 0.0}

\newcommand{\D}{\mathsf{D}}

\begin{document}
\title{  Integrated Sensing and Communication in the Finite Blocklength Regime}
\author{\IEEEauthorblockN{Homa Nikbakht$^{1}$, Mich\`ele Wigger$^2$, Shlomo Shamai (Shitz)$^3$, and H.~Vincent Poor$^1$}
	\IEEEauthorblockA{$^1$Princeton University,   $\quad ^2$LTCI,   {T}$\acute{\mbox{e}}$l$\acute{\mbox{e}}$com Paris, IP Paris,  $\quad ^3$Technion \\
		\{homa, poor\}@princeton.edu, michele.wigger@telecom-paris.fr,  sshlomo@ee.technion.ac.il}}
\maketitle

 \begin{abstract}
 A point-to-point integrated sensing and communication (ISAC) system is considered  where a transmitter conveys a message to a receiver over a discrete memoryless channel (DMC) and simultaneously estimates the state of the channel through the backscattered signals of the emitted waveform. 
  We derive achievability and converse bounds on the rate-distortion-error tradeoff in the finite blocklength regime, and also characterize the second-order rate-distortion-error region for the proposed setup. Numerical analysis shows that our proposed  joint ISAC scheme significantly outperforms traditional time-sharing based  schemes where the available resources are split between the sensing and communication tasks. 
 \end{abstract}
 \section{Introduction}
Integrating sensing capabilities into a communication network is a  promising approach to resolve the challenges of  the upcoming sixth generation (6G) wireless communication system   \cite{Liu2022, Mateos2022, Zhang2022, Liu2023, Liu2022A}. In fact,  network sensing functionality  
 is a key enabler to allow sensory data collection from the environment, which is required in applications such as industrial robots and autonomous vehicles. A recent paradigm, called \emph{integrated sensing and communication (ISAC)},  suggests to fully integrate the sensing functionality into the  communication functionality \cite{Cheng2022, Li2022, Wei2023}. In other words,  ISAC  systems  jointly perform both the sensing and communication tasks using common hardware, antenna(s) and spectrum. The benefits of such a joint approach are reductions in  hardware and signaling costs and  improvements in energy consumption and spectral efficiency   \cite{Tan2021, Chaccour2023}.  
 
 Despite a considerable amount of  interesting ISAC research efforts, the fundamental performance limits, and thus the inherent  tradeoffs between sensing and communication performances of optimal systems, remain unsolved. In particular, while 
\cite{An2023, Joudeh2022, Kobayashi2018, Kobayashi2019, Ahmadipour2022, AhmadipourISIT, AhmadipourJSAIT, Hua2023, Yao2023} determined the information-theoretic  fundamental performance limits for the asymptotic infinite blocklength regime, the focus of this article lies on  the performances of real codes at finite blocklengths. 

Specifically, in this work we consider a  point-to-point ISAC system in which the transmitter conveys a message to a receiver over a discrete memoryless state-dependent channel, and in addition, based on a generalized feedback signal, it estimates the memoryless state sequence of the channel so as to minimize a given distortion criterion. 
 We derive achievability and converse bounds on the optimal tradeoff between the communication rate and decoding error and the sensing distortion. Our  achievability and converse bounds are close, and coincide up to third-order terms in the asymptotic regimes of infinite blocklengths. For this asymptotic regime we thus refine the capacity-distortion result in \cite{Kobayashi2018,Ahmadipour2022} to the optimal scaling of the rate as a function of the allowed distortion and decoding  error probability.  
The finite-blocklength behavior of ISAC has already  been studied in \cite{Shen2023}, however for a Gaussian channel  model  where a single state (the channel coefficient) governs the entire transmission and the receiver wishes to estimate this state with smallest possible squared-error. In our setup, the state is described by a memoryless sequence impacting the various channel uses and the goal of the estimation is to reconstruct this sequence with minimum distortion.

\section{Problem Setup} \label{sec:setup}

\begin{figure}[t]
%\footnotesize
  \centering
\begin{tikzpicture}[scale=1, >=stealth]
\centering
\footnotesize
\tikzstyle{every node}=[draw,shape=circle, node distance=0.5cm];
%\draw (0,0) rectangle (1,1);
%\node[draw =none] at (0.5,0.5) {Tx};
\draw [thick, ->] (0,0.5)--(1,0.5);
\node[draw =none] at (0.5,0.65) {$M$};
\draw (1,0) rectangle (3,1);
\node[draw =none] at (2,0.5) {$f_i^{(n)}(M, Z^{i-1})$};
\draw [thick, ->] (3,0.5)--(3.9,0.5);
\node[draw =none] at (3.5,0.7) {$X_i$};
\draw (3.9,0) rectangle (6.1,1);
\node[draw =none] at (5,0.5) {\footnotesize $W(y_i,z_i|x_i,s_i)$};
\draw [thick, ->] (6.1,0.5)--(6.8,0.5);
\node[draw =none] at (6.4,0.7) {$Y_i$};
%\node[draw =none] at (6.5,0.7) {$\hat {\mathsf H}$};
\draw (6.8,0) rectangle (8.2,1);
\node[draw =none] at (7.5,0.5) {$g^{(n)}(Y^n)$};
\draw [thick, ->] (8.2,0.5)--(8.7,0.5);
%\draw [dashed] (4.4,-0.1) rectangle (9.1,1.1);
\node[draw =none] at (8.4,0.75) {$\hat M$};
\draw [thick, ->] (5,-0.5)--(5,0);
\node[draw =none] at (5,-0.7) {$S^n$};
\draw [thick, ->] (5,1)--(5,2.5)--(3,2.5);
\node[draw =none] at (2,2.5) {$h(X^{n}, Z^{n})$};
\draw (1,2) rectangle (3,3);
\draw [thick, ->] (1,2.5)--(0,2.5);
\node[draw =none] at (0.5,2.75) {$\hat S^n$};
\node[draw =none] at (5.4,1.2) {$Z_{i-1}$};
\draw [thick, ->] (5,1.5)--(2,1.5);
%\draw [thick, ->] (4.55,0.5)--(4.55,2.25)--(4,2.25);
%\draw [thick, ->, blue] (5.6,-0.3)--(7.7,-0.3)--(7.7,0);
%\node[draw =none] at (8.75,-0.5) {${\color{blue}S_{1,i}}$};
\draw [dashed] (0.8,-0.5) rectangle (3.2,3.5);
\node[draw =none] at (2,-0.8) {Transmitter};
\draw[thick, ->] (2,2)--(2,1);
\end{tikzpicture}
\caption{ISAC System model.}
\label{fig1}
  \end{figure}
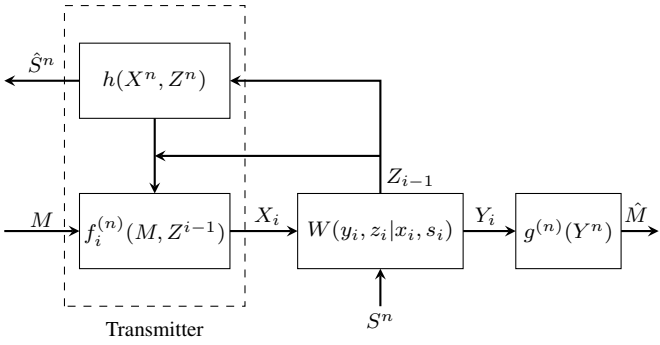

Consider the point-to-point setup in Figure~\ref{fig1} where a transmitter wishes to communicate a message $M$, which is uniformly distributed over a set $\{1, \ldots, \mathsf{M}\}$, to a receiver over  a state-dependent memoryless channel and at the same time wishes to estimate the channel state sequence based on a  generalized feedback signal.  We consider the  discrete memoryless state-dependent channel  with finite input alphabet $\mathcal X$, finite channel state alphabet $\mathcal S$, finite feedback alphabet $\mathcal Z$, finite output alphabet $\mathcal Y$ and the channel transition law
\begin{equation}
P_{Y^n Z^n |X^n S^n} (y^n, z^n | x^n, s^n ) = \prod_{i = 1}^ n W( y_i, z_i |x_i, s_i)\end{equation}
for a given conditional pmf $W(\cdot, \cdot| \cdot, \cdot)$.

So, if $M=m$,  at a given time $i \in \{1, \ldots, n\}$ and after observing the feedback sequence $Z_{i-1}$, the transmitter sends an input symbol \begin{equation}
X_i = f_i^{(n)}(m, Z^{i-1})
\end{equation}
where  for any $ i \in \{1, \ldots, n\}$  the encoding function $f_i^{(n)}$ is defined on appropriate domains. 
The transmitter also estimates the channel state $S^n$  that is  i.i.d according to a given distribution $P_S$  as 
\begin{equation}\label{eq:estimation}
\hat S^n = h^{(n)}(Z^n, X^n),
\end{equation}
based on  a block-estimation function $h^{(n)}$, defined on appropriate domains.

After observing the channel outputs $Y^n$,  the receiver decodes the message $M$ as
\begin{IEEEeqnarray}{rCl}
\hat M &=& g^{(n)}(Y^n) ,
\end{IEEEeqnarray} 
where $g^{(n)}$ is a decoding function on appropriate domains. 
The quality of the state estimation at the transmitter is measured by the expected average per-block distortion
\begin{IEEEeqnarray}{rCl}
\Delta^{(n)} : = \mathbb E [d(S^n, \hat S^n) ] &=& \frac{1}{n} \sum_{i = 1}^n \mathbb E [d(S_{i}, \hat S_{i})] 
\end{IEEEeqnarray}  
  for a given bounded per-symbol  distortion function $d(\cdot, \cdot)$.

The decoding error probability is defined as:
\begin{equation} \label{eq:6}
\epsilon^{(n)} : = \Pr [ \hat M \neq M]. 
\end{equation}

\begin{definition}
Given a blocklength $n$, the rate-distortion-error triple $(\mathsf{R}, \D, \epsilon)$ is  said  to be achievable, if there exist encoding, decoding, and estimation functions $\{ f^{(n)}, g^{(n)}, h^{(n)}\}$  satisfying
\begin{IEEEeqnarray}{rCl}
\frac{1}{n} \log_2 (\mathsf{M}) & \geq & \mathsf{R},\\
 \epsilon^{(n)} &\le & \epsilon, \label{eq:epsilon}\\
 \Delta^{(n)} & \le & \D.
%\lim_{n \to \infty} \Delta_r^{(n)} & \le & \Dr. 
\end{IEEEeqnarray}
%The capacity-distortion trade-off $\mathsf C (\D)$ is the largest rate $\\mathsf{R}$ such that the rate-distortion tuple $(\\mathsf{R}, \D)$ is achievable.
\end{definition}

\section{Optimal Estimator} 
%\subsection{Estimating}
For the described memoryless setup, the optimal state estimator is a symbolwise estimator  applied to the transmitter's observations $X^n$ and $Z^n$:
\begin{IEEEeqnarray}{rCl} \label{eq:10}
\hat{ S}^n= [(\hat s^*(X_1, Z_1), \hat s^*(X_2, Z_2), \ldots, \hat s^*(X_n, Z_n) ],
\end{IEEEeqnarray}
where 
\begin{IEEEeqnarray}{rCl}
\hat s^*(x, z) : = \text{arg}\min_{s" \in \hat{\mathcal S}} \sum_{s \in \mathcal S} P_{S|XZ} (s|x,z) d(s, s'),
\end{IEEEeqnarray}
with 
\begin{IEEEeqnarray}{rCl}
P_{S|XZ} (s|x,z) = \frac{P_S(s) P_{Z|SX}(z|s,x)}{\sum_{\tilde s \in \mathcal S} P_S(\tilde s) P_{Z|SX}(z|\tilde s, x)}. 
\end{IEEEeqnarray}
The proof of optimality  of this symbolwise estimator relies on the Markov chain relation \begin{IEEEeqnarray}{rCl}
(X^{i-1}, X_{i+1}^n, Z^{i-1}, Z_{i+1}^n) \multimap \hspace{-0.15cm}- (X_i,Z_i) \multimap \hspace{-0.15cm}- S_i,
\end{IEEEeqnarray} 
see  \cite[Appendix A]{Ahmadipour2022}  for more details.
%
%The estimation cost of such an estimator is thus equal to 
%\begin{IEEEeqnarray}{rCl} \label{eq:13}
%c(x) : = \mathbb E \left [ d (S,  \hat s^*(X, Z) ) | X = x\right ].
%\end{IEEEeqnarray}

\section{Main Results}
Given two random variables $X$ and $Y$ having joint probability mass function (pmf) $P_{XY}(x,y)$, define their information density 
\begin{IEEEeqnarray}{rCl}
i(X;Y): = \log \frac{P_{Y|X}(y|x)}{P_Y(y)},
\end{IEEEeqnarray}
and notice that the expectation of the information density equals the mutual information $I(X;Y)= \mathbb E [i(X, Y)]$. Denote the  higher central moments of the information density as
%\begin{subequations}
\begin{IEEEeqnarray}{rCl}
%E&:=& \mathbb E [i(X, Y)] \notag \\
%&  =& \sum_{x  \in \mathcal X} \sum_{y \in \mathcal Y} P_{X} (x) P_{Y |X} (y|x)\log  \frac{ P_{Y |X} (y|x)}{P_{Y} (y)},\label{eq:E} \\
\mathsf{V} &: = & \text{Var} [i(X;Y)] \notag \\
& = & \sum_{x , y } P_{X} (x) P_{Y | X} (y| x)\log^2 \frac{ P_{Y | X} ( y| x)}{P_{ Y} ( y)} - I(X;Y)^2,\IEEEeqnarraynumspace \label{eq:V} \\
\mathsf{T} &:=& \mathbb E [|i(X; Y) - \mathbb  I( X; Y) |^3] \notag \\
& = & \sum_{x, y} P_{X} (x) P_{ Y | X} ( y| x) \left |\log  \frac{ P_{ Y | X} (y| x)}{P_{ Y} ( y)} - I(X;Y) \right|^3. \label{eq:T}
\end{IEEEeqnarray}
%\end{subequations}

Our main results are the following theorems on the rate-distortion-error tradeoff. 
\begin{theorem}[Achievability Bound]\label{th1}
Given a blocklength $n$, the rate-distortion-error tradeoff $(\mathsf{R}, \D, \epsilon)$ is achievable if there exists a $P_X$ and a constant $ \mathsf{K} >0$ such that the following two conditions are satisfied,
\begin{IEEEeqnarray}{rCl}
\mathsf{R} & \le&I( X;  Y) - \sqrt{\frac{\mathsf{V}}{n}} \mathbb Q^{-1} \left (\epsilon - \beta_u \right) -  \mathsf{K}\frac{\log (n)}{n},\label{eq:Ru}\IEEEeqnarraynumspace \\
\D & \ge & \sum_{x \in \mathcal X} \sum_{ s \in \mathcal S} \sum_{z \in \mathcal Z} d (s, \hat s^*(x, z)) P_X(x) P_S(s) P_{Z|XS}(z|x,s),  \label{eq:D}\IEEEeqnarraynumspace
\end{IEEEeqnarray} 
with 
\begin{equation}\label{eq:betau}
\beta_u := \frac{1}{n^{ \mathsf{K}}} + \frac{0.7975 \mathsf{T}}{\sqrt{n \mathsf{V}^3}},
\end{equation}
and where the mutual information $I(X;Y)$ and  the two central moments $\mathsf{V}$ and $\mathsf{T}$ are defined based on the joint pmf $P_{XY}(x,y)=P_X(x)P_{Y|X}(y|x)$.
\end{theorem}
\begin{IEEEproof}
See Section~\ref{sec:achiv}.
\end{IEEEproof}

We also have the following converse bound.
\begin{theorem}[Converse Bound]\label{th2}
Given the blocklength $n$, a rate-distortion-error triple $(\mathsf{R}, \D, \epsilon)$ is not achievable if for all $\delta > 0$ and pmfs $P_X$ satisfying \eqref{eq:D} the following lower bound holds: 
\begin{IEEEeqnarray}{rCl}
\mathsf{R} & \ge&I( X;  Y) - \sqrt{\frac{\mathsf{V}}{n}} \mathbb Q^{-1} \left (\epsilon + \beta_l \right) + \frac{\log(n)}{2n} - \frac{\log \delta}{n},  \label{eq:Rl}\IEEEeqnarraynumspace 
\end{IEEEeqnarray} 
where
\begin{equation}\label{eq:betal}
\beta_l : = \frac{0.7975\mathsf{T}}{\sqrt{n \mathsf{V}^3}} + \frac{\delta}{\sqrt{n}}.
\end{equation}
\end{theorem}
\begin{IEEEproof}
The proof of the bound in \eqref{eq:Rl} follows similar steps as  the proof of \cite[Lemma 58]{Yuri2012}, where one has to integrate the optimal estimator in \eqref{eq:10}. See Appendix~\ref{AppA} for details. \end{IEEEproof}

\begin{proposition} \label{prop1}
Given $\D$, $\epsilon$ and large blocklengths $n$, the largest rate $\mathsf R$ such that the triple $(\mathsf R,\D, \epsilon)$ is achievable, is given by
%For sufficiently large $n$,   the rate-distortion-error triple $(\mathsf{R}, \D, \epsilon)$ is achievable if, and only if,  there is a $P_X$ such that condition \eqref{eq:D} holds  and 
\begin{equation}
\mathsf R_{\max}(\D, \epsilon,n)= \max_{P_X} \left [I(X; Y) - \sqrt{\frac{\mathsf{V}}{n}} \mathbb Q^{-1} \left (\epsilon\right)  + O\left (\frac{\log n}{n}\right)\right ], \label{eq:R2}
\end{equation}
where the maximum is over all pmfs $P_X$ satisfying \eqref{eq:D}.
\end{proposition}
\begin{IEEEproof} 
By the differentiability $\mathbb Q^{-1}$ and by the forms of $\beta_u$ and $\beta_l$ in \eqref{eq:betau} and \eqref{eq:betal}, we have
\begin{IEEEeqnarray}{rCl}\label{eq:24n}
\mathbb Q^{-1} (\epsilon - \beta_u) &=& \mathbb Q^{-1}(\epsilon) + O\left (\frac{1}{\sqrt{n}}\right),  \\
\mathbb Q^{-1} (\epsilon + \beta_l) &=& \mathbb Q^{-1}(\epsilon) + O\left (\frac{1}{\sqrt{n}}\right). \label{eq:25n}
\end{IEEEeqnarray}
Substituting \eqref{eq:24n} into  \eqref{eq:Ru}, and \eqref{eq:25n} into \eqref{eq:Rl}  proves the proposition.
\end{IEEEproof}

\begin{remark}
Equality \eqref{eq:R2} agrees with \cite[Theorem 49]{Yuri2012} which determines the second-order coding rate of a DMC in the finite blocklength regime. \end{remark}

\section{Comparisons and Examples}
In this section, we evaluate  Theorems~\ref{th1} and \ref{th2} numerically for a binary example and compare them also with the performance of two baseline schemes that are frequently employed in practice.
\subsection{Time-Sharing Schemes}
Many practical systems employ a basic resource-sharing approach where a fraction of the resources (here $(1-\gamma) n$ channel uses) are dedicated only to the communication task and the remaining resources (here $\gamma n$ channel uses) to the sensing task, each one completely ignoring the other task. A slightly improved scheme uses the resources for the communication task also for some basic sensing, but using the waveform that is best for communication, and similarly uses the resources for the sensing task also for communication, but using the best waveform for sensing.

\subsubsection{Basic Resource-Sharing Scheme}
Given  time-sharing parameter $\gamma \in[0,1]$, the performance of the basic resource-sharing scheme described above  achieves rate
\begin{equation}
\mathsf{R}= (1-\gamma) \mathsf R_{\max}
\end{equation}
and distortion
\begin{equation}
\D= \gamma \D_{\min}+(1-\gamma)\D_{\textnormal{trivial}}, 
\end{equation}
where $\mathsf R_{\max}$ is the largest achievable rate:
\begin{equation}\label{eq:Rmax}
\mathsf R_{\max} := \max_{P_X} \left[ I( X;  Y) - \sqrt{\frac{\mathsf{V}}{n}} \mathbb Q^{-1} \left (\epsilon - \beta_u \right) -  \mathsf{K}\frac{\log (n)}{n} \right]
\end{equation} 
%Given $\epsilon$ and $n$, the following two extreme rate-distortion points are achieved under this scheme:
%\begin{IEEEeqnarray}{rCl}
%\left ( (\mathsf{R} = 0, \D = \D_{\min}), (\mathsf{R} = \max \text{RHS (Eq. \eqref{eq:Ru})}, \D_{\text{Trivial}})\right), \IEEEeqnarraynumspace
%\end{IEEEeqnarray}
%where 
and $\D_{\min}$ denotes the best possible distortion while $\D_{\text{trivial}}$ denotes the distortion achieved by the optimal  trivial estimator that does not exploit the feedback: 
\begin{IEEEeqnarray}{rCl}\label{eq:Dmin}
\D_{\min} &: =& \min_{P_X} \sum_{x \in \mathcal X} \sum_{s \in \mathcal S} \sum_{z \in \mathcal{Z}} P_X(x) P_S(s) P_{Z|SX}(z|s,x) \hat{s}^*(x,z), \nonumber\\\label{eq:Dmin}\\
\D_{\text{trivial}} & :=& \min_{s' \in \mathcal S} \sum_{s \in \mathcal S} P_S(s) d(s, s').
\end{IEEEeqnarray}

\subsubsection{Improved Resource-Sharing Scheme} %Reconsider the basic time-sharing scheme from the previous suubsection, but modify it so that during the $\gamma$-fraction of the channel uses dedicated to sensing, the emitted waveform (with the optimal $P_X$ according to \eqref{eq:Dmin}) is  also used to  communicate data to the receiver, and similarly, during the $(1-\gamma)$-fraction of the channel uses dedicated to communication, the feedback from the emitted waveform (with the optimal $P_X$ according to \eqref{eq:Rmax}) is also used  for sensing purposes. 

For a given time-sharing parameter $\gamma \in[0,1]$, the improved resource-sharing scheme  achieves rate
\begin{equation}
\mathsf{R}= \gamma \mathsf{R}_{\textnormal{sense}}+(1-\gamma) \mathsf{R}_{\max} 
\end{equation}
and distortion
\begin{equation}
\D= \gamma \D_{\min}+(1-\gamma)\D_{\textnormal{comm}}, 
\end{equation}
where 
\begin{IEEEeqnarray}{rCl}
\D_{\textnormal{comm}}: = \sum_{x \in \mathcal X} P_{X}^\star (x)  \sum_{s \in \mathcal S} \sum_{z \in \mathcal{Z}}P_S(s) P_{Z|SX}(z|s,x) \hat{s}^*(x,z) \IEEEeqnarraynumspace
\end{IEEEeqnarray}
for $P_{X}^\star$ the optimizer in \eqref{eq:Rmax} 
and 
\begin{IEEEeqnarray}{rCl}
\mathsf{R}_{\textnormal{sense}}:=  I( X;  Y) - \sqrt{\frac{\mathsf{V}}{n}} \mathbb Q^{-1} \left (\epsilon - \beta_u \right) -  \mathsf{K}\frac{\log (n)}{n} 
\end{IEEEeqnarray}
evaluated for $P_{XY}=P_X'P_{Y|X}$ with $P_X'$ the optimizer of \eqref{eq:Dmin}.

%followed by some numerical analysis. % are presented in Section~\ref{sec:C}. %We start by the following definitions. 
%\begin{IEEEeqnarray}{rCl}
%P_{X,\max} &\triangleq& \text{arg} \max _{P_x} I(X;Y), \quad \D_{\max} \triangleq \sum_{x \in \mathcal X} P_{X, \max} c(x), \IEEEeqnarraynumspace \\
%P_{X, \max}^{(u)} &\triangleq& \text{arg} \max _{P_x} \text{RHS (\eqref{eq:Ru})}, \;  \D_{\max}^{(u)} \triangleq \sum_{x \in \mathcal X} P_{X, \max}^{(u)}c(x),\\
%P_{X, \max}^{(l)} &\triangleq& \text{arg} \max _{P_x} \text{RHS (\eqref{eq:Rl})}, \;  \D_{\max}^{(l)} \triangleq \sum_{x \in \mathcal X} P_{X, \max}^{(l)}c(x).
%\D_{\min} &\triangleq & \min_{P_X} \sum_{x \in \mathcal X} P_X(x) c(x), \; \D_{\text{Trivial}}  \triangleq \min_{s' \in \mathcal S} \sum_{s \in \mathcal S} P_S(s) d(s, s').
%\end{IEEEeqnarray}
\subsection{Binary Channel with Multiplicative Bernoulli State}
Consider the channel 
\begin{equation} \label{eq:channel}
Y = SX,
\end{equation}
with binary alphabets $\mathcal X = \mathcal S = \mathcal Y \in \{0,1\}$ and where the state is Bernoulli-$q$ with $q \in (0,1)$ and the feedback is perfect, i.e., $Z = Y$. We consider the Hamming distortion measure $d(s, \hat s) = s \oplus \hat s$.  

To compare  the performance specified in Theorems~\ref{th1} and \ref{th2} with each other and with the performance of the two baseline time-sharing schemes, we parametrize the binary input distribution $P_X$ by $\alpha := \Pr [X = 1]$. We also notice that the channel in \eqref{eq:channel} is equivalent to a Z-Channel: input $0$  always leads to the output symbol 0 and input $1$ leads to output 0 with probability $1-q$ and to output 1  with probability $q$. The mutual information between input and output of the channel is then obtained as
\begin{subequations}\label{eq:22}
\begin{IEEEeqnarray}{rCl}
 I( X; Y) &=& H_b(q \alpha) - \alpha H_b(q),
 \end{IEEEeqnarray}
 where $H_b(x) = - x\log (x) -(1-x)\log(1-x)$ is the binary entropy function. 
 For the second and third central moments of the information density we have
 \begin{IEEEeqnarray}{rCl}
\mathsf{V}_\alpha &=& \alpha \left ( q \log^2 \frac{1}{\alpha} + (1- q)\log^2\frac{1-q}{1-q\alpha}\right) \notag \\
&&+ (1-\alpha) \log^2 \frac{1}{1- q\alpha} -  I( X; Y) ^2, \\
\mathsf{T}_\alpha &=& \alpha q \left | \log \frac{1}{\alpha} - I( X; Y)  \right|^3\nonumber \\
&& + (1- q)\left |\log \frac{1-q}{1-q\alpha} -  I( X; Y) \right|^3 \notag \\
&&  + (1-\alpha) \left |\log \frac{1}{1- q\alpha} -  I( X; Y) \right|^3,
\end{IEEEeqnarray}
\end{subequations}

We can then substitute $ I( X; Y) $ and $\mathsf{V},\mathsf{T}$ from \eqref{eq:22} into \eqref{eq:Ru} and \eqref{eq:Rl} to obtain the desired bounds on the rate. %one can calculate the  achievability and converse bounds on the transmission rate of  this channel under the distortion constraint \eqref{eq:D}.

 To calculate the distortion bound \eqref{eq:D}, notice that whenever $x = 1$, then $z = y = s$ and thus the distortion is zero. On the other hand, when $x=0$ then $y=0$ and the transmitter does not receive any information about the state of the channel.  In this case, the optimal estimator is to choose the most likely state symbol, i.e. $\hat{s}=0$ if $q<1/2$ and $\hat{s}=1$ if $q\geq 1/2$. We combine these observations to obtain the following bound: 
\begin{IEEEeqnarray}{rCl}
\D &\ge& P_X(0)\sum_{s, y} d(s, \hat s^*(x = 0, y) )P_S(s) P_{Y|XS}(y|x = 0, s) \IEEEeqnarraynumspace\\
&=&P_X(0) \sum_{s \in \mathcal S}  d(s, \hat s^*(x = 0, y= 0) )P_S(s) \\
& = & (1- \alpha)\min \{q, 1-q\}.
\end{IEEEeqnarray}
In other words, a distortion constraint imposes the following bound on $\alpha$: 
\begin{equation}
\alpha \geq 1-   \frac{\D}{\min \{q, 1-q\}}.
\end{equation}

Thus, for this example Theorem~\ref{th1} states that for any $\D >0$, all triples $(\mathsf{R}, \D, \epsilon)$  are achievable if
\begin{equation}\label{eq:Ralpha}
\mathsf{R} \leq  \max_{\alpha,  \mathsf{K}\geq 0} I( X;  Y) - \sqrt{\frac{\mathsf{V}}{n}} \mathbb Q^{-1} \left (\epsilon - \beta_u \right) -  \mathsf{K}\frac{\log (n)}{n} ,
\end{equation}where the maximization is over all $\alpha\in[0,1]$ satisfying $1 \geq \alpha \geq  1-   \frac{\D}{\min \{q, 1-q\}}$. Theorem~\ref{th2} states that for any $\D>0$ all triples $(\mathsf{R}, \D, \epsilon)$ satisfying 
\begin{equation}\label{eq:R_conv}
\mathsf{R} \geq  \max_{\alpha,\delta>0} I( X;  Y) - \sqrt{\frac{\mathsf{V}}{n}} \mathbb Q^{-1} \left (\epsilon + \beta_l \right) + \frac{\log(n)}{2n} - \frac{\log \delta}{n} 
\end{equation} 
are not achievable. Here, the maximization is again over values  $\alpha\in \left[ \min\left\{0,  1-   \frac{\D}{\min \{q, 1-q\}} \right\}, 1 \right]  $.

Notice that for this channel   (which is a Z-channel)  the capacity is equal to \cite{Tallini2002}
\begin{IEEEeqnarray}{rCl}
 \mathsf{C} = \log (1 + q (1-q)^{\frac{1-q}{q}}),
\end{IEEEeqnarray}
and is achieved for 
\begin{equation}
P_X^\star (1)=\alpha^\star = \frac{1}{q\left(1+2^{\frac{H_b(q)}{q}}\right)}.
\end{equation}
The distortion achieved with this capacity-achieving $\alpha^\star$ is $\D_{\textnormal{comm}}= (1-\alpha^\star) \min\{q, 1-q\}$.

\subsection{Numerical Analysis}
\begin{figure}[t]
\center
				\includegraphics[width=0.48\textwidth]{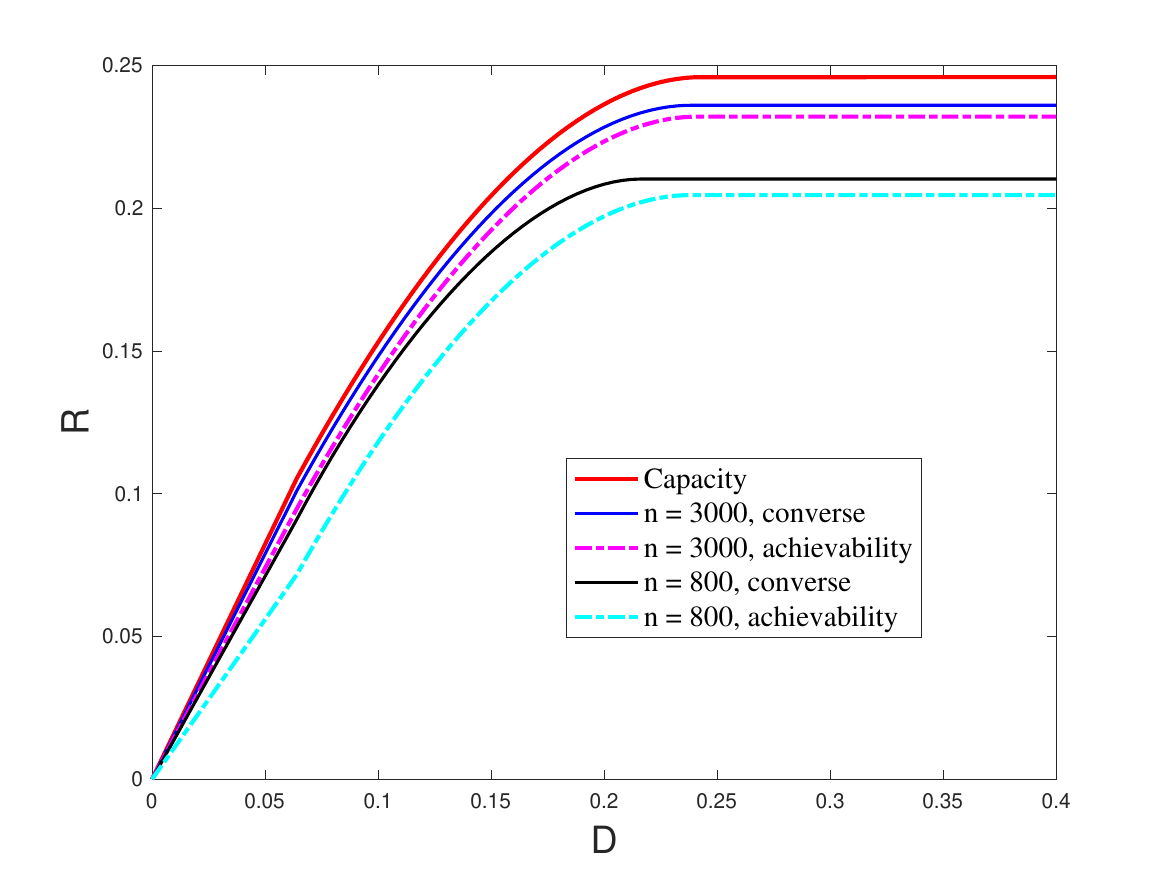}
\caption{Achievability and converse bounds on the rate-distortion-error trade-off of Theorems~\ref{th1} and \ref{th2} for $\epsilon = 0.05$, $q = 0.4$, and different values of $n$. }
\label{fig1R}
			\end{figure}

 Fig.~\ref{fig1R} illustrates the achievability and converse bounds on the rate-distortion-error tradeoff presented in \eqref{eq:Ralpha} and \eqref{eq:R_conv} for $\epsilon = 0.05$ and $q = 0.4$. As can be seen from this figure the bounds are tight for large values of $n$. Notice that for $q=0.4$ the capacity of the channel is $\mathsf{C}= 0.246$ and the achieved distortion is $\D_{\textnormal{comm}}=0.2432$.

% \begin{figure}[t]
%\center
%				\includegraphics[width=0.48\textwidth]{figISACm2.eps}
%\caption{Rate-distortion-error trade-off for $\epsilon_1 = 10^{-3}$, $q = 0.4$, $  \mathsf{K} = 0.5$ and different values of $n$. }
%\label{fig2R}
%			\end{figure}
%			
%			
			\begin{figure}[t]
\center
		\includegraphics[width=0.48\textwidth]{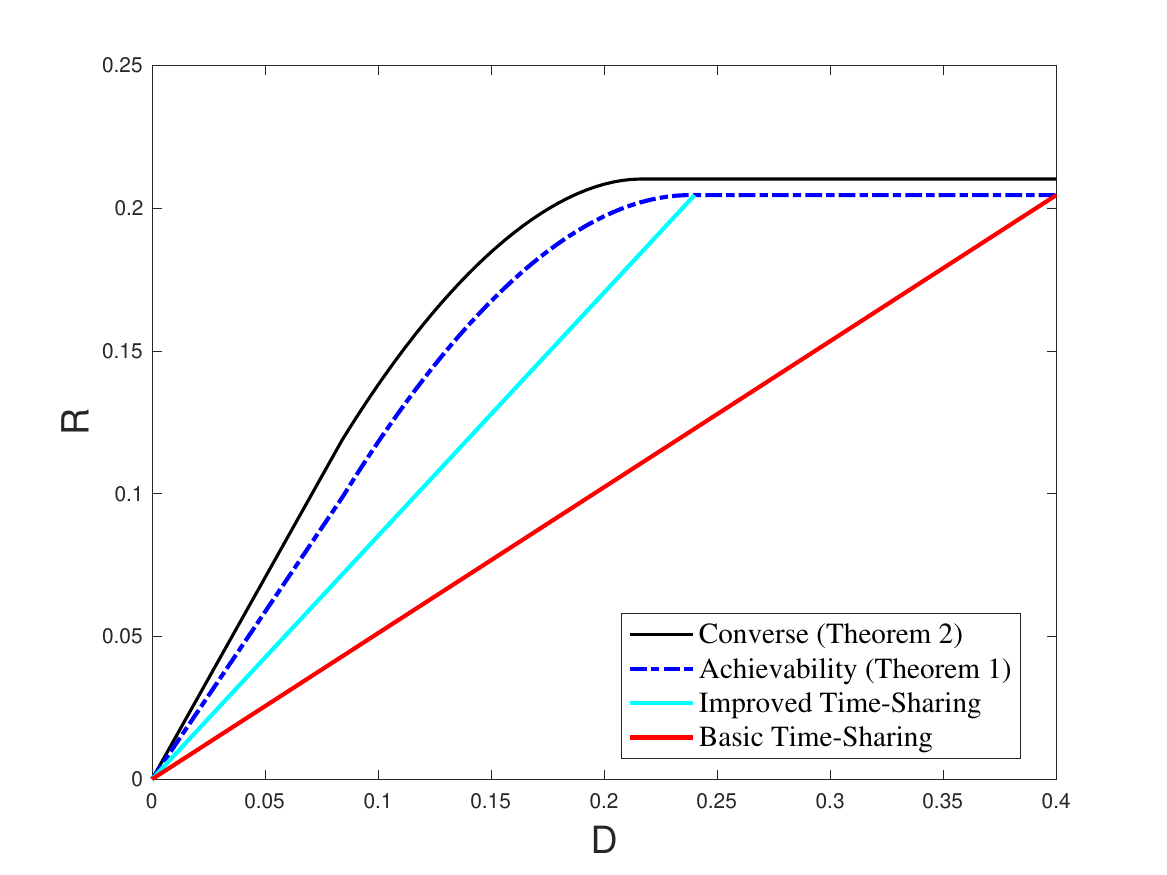}
\caption{Comparison of  the rate-distortion-error trade-off  in Theorems~\ref{th1} and \ref{th2} with the basic and improved resource-sharing schemes for $\epsilon = 0.05$, $q = 0.4$,  and $n = 700$.}
\label{fig3R}
			\end{figure}
%			In Fig. \ref{fig2R}, we study the effect of using the capacity-achieving \label{eq:alphastar}
%
%			instead of the maximizer in  \eqref{eq:Ralpha}. 
%\begin{equation}
%R_{c,n} : = \text{RHS (Eq.\eqref{eq:Ru})}, X \sim Bernoulli-\alpha^*
%\end{equation}
%whereas the solid regions are the rate-distortion tradeoff for which $\mathsf{R} \le R_{\max, n}  $ where
%\begin{equation}
%R_{\max,n} : = \max_{P_X} \text{RHS (Eq. \eqref{eq:Ru}}).
%\end{equation}
%As can be seen from this figure, for small values of $n$, the capacity-achieving distribution does not maximize the bound in \eqref{eq:Ru}. 

Fig.~\ref{fig3R} compares the rate-distortion-error tradeoff achieved by our scheme with the tradeoff achieved under the basic and improved resource-sharing schemes.  As can be seen from this figure, our scheme outperforms the other two baseline schemes.

\section{Proof of Theorem~\ref{th1}}\label{sec:achiv}
\subsection{Codebook Generation}
Choose $P_X$ satisfying \eqref{eq:D}. The codebook $\mathcal C = \{x^n(m)\}_{m= 1}^{\mathsf{M}}$   is generated by randomly and independently choosing each entry according to $P_X$. 
\subsection{Encoding}
 To send a message $m$, the transmitter encodes this message via the codeword $x^n(m)$ and sends it over the channel. 

\subsection{Estimation}
After observing the feedback sequence $Z^n = z^n$, the transmitter estimates the channel state through \eqref{eq:10}. 
\subsection{Decoding}
Given the channel outputs  $Y^n=y^n$, the receiver  estimates the message $M$ by choosing  the index $\hat{m}$ that corresponds to the codeword $x^n(\hat{m})$  that maximizes the information density:
\begin{equation}
\hat{m} := \textnormal{arg} \max_{m}  i \left( x^n(m);  y^n  \right). 
\end{equation} 
The receiver then produces the guess $\hat{M}=\hat{m}$.

\subsection{Error Analysis}
To analyze $\Pr [\hat{M} \neq M]$, we use the threshold-based metric bound in  \cite{Yuri2012}. For any  $\gamma \in \mathbb{R}$, we have
\begin{IEEEeqnarray}{rCl} \label{eq:23}
\Pr [\hat M \neq M] &\le& \Pr [i( X^n; Y^n) \le \gamma] + \mathsf{M} \cdot \Pr [i(\bar{ X}^n;  Y^n) \ge \gamma ],\IEEEeqnarraynumspace
\end{IEEEeqnarray}
where $\bar{X}^n \sim P_{X^n}$ and is independent of $X^n$ and $Y^n$.  
 We will set 
 \begin{IEEEeqnarray}{rCl}\label{eq:gamma}
%\tilde K &=& \frac{\gamma - nE}{\sqrt{n \mathsf{V}}} \\
\gamma &: = & \log \mathsf{M} + \mathsf{K} \log n,
\end{IEEEeqnarray}
 for some $\mathsf{K}>0$, and 
  employ the Berry-Esseen theorem  and Bayes' formula  to evaluate the two terms on the right-hand side of \eqref{eq:23}.

By the strengthening of the Berry-Esseen theorem in  \cite{Beeck1972},  and because $\mathbb{E}[ i(X^n;Y^n)] = n I(X;Y)$, we have with the definition in \eqref{eq:gamma}
\begin{IEEEeqnarray}{rCl}
\Pr\left[i( X^n; Y^n) \leq  \gamma\right] &\leq& \mathbb{Q}\left (\frac{-\log \mathsf{M} + nI(X;Y) -  \mathsf{K} \log (n)}{\sqrt{n \mathsf{V}}}\right)   \nonumber \\
&&+ \frac{0.7975 \mathsf{T}}{\sqrt{n \mathsf{V}^3}}.\label{eq:44}
\end{IEEEeqnarray}

To bound $\Pr [i(\bar{ X}^n;  Y^n) \ge \gamma]$, we first use Bayes' formula to write
\begin{IEEEeqnarray}{rCl}
 	P_{{X}^n}(\overline{ x}^n)  & =&  \frac{P_{{Y}^n}({y^n})P_{{X}^n|{Y}^n}(\overline{x}^n|{y}^n)}{P_{{Y}^n|{X}^n}({y}^n|\overline{x}^n)}  \\
	&=& P_{\bar{X}^n|{Y}^n}(\overline{x}^n|{y}^n)2^{-i(\bar{x}^n;  y^n)}.
\end{IEEEeqnarray}
For any ${y}^n \in \mathbb{R}^n$, we then have 
 \begin{align}
 	&\sum_{\bar x^n \in \mathcal X} \mathbbm{1}\left\{i(\bar{x}^n; y^n) > \gamma\right\}P_{{X}^n}(\overline{x}^n)\notag\\
 	&= \sum_{\bar x^n \in \mathcal X} 2^{-i(\bar{x}^n; y^n)}\mathbbm{1}\left\{\frac{  P_{ Y^n | X^n } (y^n | \overline{x}^n) }{P_{Y^n }(y^n )} > 2^{\gamma}\right\}  \nonumber \\
	& \hspace{3cm} \cdot P_{{X}^n|{Y}^n}(\overline{x}^n|{ y}^n)\\[1ex]
 	&\leq \sum_{\bar x^n \in \mathcal X} 2^{-i(\bar{x}^n; y^n)}\frac{  P_{ Y^n |  X^n} (y^n| \overline{x}^n) }{P_{Y^n }( y^n)}2^{-\gamma} P_{{ X^n}|{Y^n}}(\overline{ x}^n|{ y}^n)\notag\\
 	&= \sum_{\bar x^n \in \mathcal X} P_{{X}^n|{Y}^n}(\overline{x}^n|{y}^n)2^{-\gamma} \notag\\
 	&= 2^{-\gamma}. \label{eq:106}
 \end{align}
As a consequence,
 \begin{align} 
	\Pr [i(\bar{X}^n; Y^n)  \ge \gamma] \leq 2^{-\gamma}
 \end{align}
and 
 \begin{align}\label{eq:48}
 \mathsf{M} \Pr [i(\bar{ X}^n;  Y^n) \ge \gamma ] \leq 2^{-\gamma + \log \mathsf{M}} = n^{-\mathsf{K}}.
 \end{align}
Combining \eqref{eq:23}, \eqref{eq:44}, and  \eqref{eq:48}, we obtain
\begin{IEEEeqnarray}{rCl}
\Pr [\hat M \neq M]  \le \mathbb{Q}\left (\frac{-\log \mathsf{M} + nI(X;Y) -  \mathsf{K} \log (n)}{\sqrt{n \mathsf{V}}}\right)  + \beta_u,  \nonumber\\
\end{IEEEeqnarray}
where $\beta_u$ is defined in \eqref{eq:betau}. 

Thus, the probability of error stays below $\epsilon$ whenever 
\begin{IEEEeqnarray}{rCl}
\epsilon - \beta_u \ge \mathbb{Q}\left (\frac{-\log \mathsf{M} + nI(X;Y) - \mathsf{K} \log (n)}{\sqrt{n \mathsf{V}}}\right),
\end{IEEEeqnarray}
or equivalently when
\begin{IEEEeqnarray}{rCl}
\log \mathsf{M} \le n I(X;Y) - \sqrt{n \mathsf{V}} \mathbb Q^{-1} (\epsilon -\beta_u) -  \mathsf{K} \log (n),
\end{IEEEeqnarray}
establishing the bound in \eqref{eq:Ru}.

\subsection{Expected Distortion}
The expected distortion can be written as
\begin{IEEEeqnarray}{rCl}
\Delta^{(n)}   &=& \frac{1}{n} \sum_{i = 1}^n \mathbb E [d(S_i, \hat S_i)] \\
& = & \sum_{x \in \mathcal X} \sum_{ s \in \mathcal S} \sum_{z \in \mathcal Z} d (s, \hat s^*(x, z)) P_X(x) P_S(s) P_{Z|XS}(z|x,s).  \nonumber \\
\end{IEEEeqnarray}
By our choice of $P_X$, our scheme thus satisfies the requirement on the distortion. 

\section{Conclusions}
We have studied the rate-distortion-error tradeoff of a  point-to-point ISAC system  where a transmitter conveys a message to a receiver over a discrete memoryless state-dependent channel  and simultaneously estimates the state of the channel.  We have derived achievability and converse bounds on the rate-distortion-error tradeoff in the finite blocklength regime.  We also have characterized the second-order rate-distortion-error region of the proposed setup. Our numerical analysis  shows that our joint design scheme significantly outperforms the resource-sharing baseline  schemes where the available resources are split between the sensing and communication tasks. In our model the receiver has no state-information. The generality of our model allows however to obtain results for perfect or partial state-information as special cases from our Theorems~\ref{th1} and \ref{th2}, simply by including the state-information as part of the receiver's output. An interesting line of future work is to study the ISAC problem with general state and channel distribution in the finite blocklength regime \cite{Chen2023}.

\section*{Acknowledgment}
The work of H. Nikbakht and H. V. Poor has been supported by the U.S National Science Foundation under Grant CNS-2128448. The work of S. Shamai (Shitz) has been supported by the US-Israel Binational Science Foundation (BSF) under grant BSF-2018710 and by the German Research Foundation (DFG) via the German-Israeli Project Cooperation (DIP), under Project SH 1937/1-1.

%the U.S National Science Foundation under Grant CNS-2128448.

\appendices 
\section{Proof of Theorem~\ref{th2}}\label{AppA}
The proof of the bound in \eqref{eq:Rl} follows similar steps as  the proof of \cite[Lemma 58, Theorem 28]{Yuri2012}. In the following, we sketch the proof of the converse bound \eqref{eq:Rl}.

Consider a random variable $Y$ on $\mathcal Y$ which can take probability measures $P_{Y|X }$ and $P_Y$. Define by $P_{Z|Y}: \mathcal Y \to \{0,1\}$ a randomized test between those two distributions where $0$ indicates that the test chooses $P_Y$.  Let  $\beta_{\alpha} (P_{Y|X}, P_Y)$ be the minimum probability of error under hypothesis $P_Y$ if the probability of error under hypothesis $P_{Y|X}$ is below $1-\alpha$. I.e., \begin{IEEEeqnarray}{rCl}
\lefteqn{\beta_{\alpha} (P_{Y|X}, P_Y) } \notag \\
&=& \min_{P_{Z|Y}: \sum_{y \in \mathcal Y} P_Y(y) P_{Z|Y}(1|y) \ge \alpha} \sum_{y \in \mathcal Y} P_Y(y) P_{Z|Y}(1|y).
\end{IEEEeqnarray} 
It is known that $\beta_{\alpha} (P_{Y|X}, P_Y)$ is the  best performance achievable among such randomized tests. 

 %As a function of $\alpha$, $\beta_{\alpha} (P_{Y|X}, P_Y)$  is a piecewise-linear convex function joining the points 
%\begin{equation}
%\alpha = \Pr [\frac{}{}]
%\end{equation}
It is easy to show that for any $\tilde \gamma>0$,
\begin{IEEEeqnarray}{rCl}
\alpha \le \Pr \left [\frac{dP_{Y|X}}{dP_Y} \ge \tilde \gamma \right ] + \tilde \gamma \beta_{\alpha} (P_{Y|X}, P_Y).
\end{IEEEeqnarray}
Equivalently, for any $\gamma_n > 0$
\begin{IEEEeqnarray}{rCl} \label{eq:beta}
\beta_{\alpha} (P_{Y^n|X^n}, P_{Y^n}) \ge \frac{1}{\gamma_n} \left ( \alpha - \Pr \left [\log \frac{dP_{Y^n|X^n}}{dP_{Y^n}} \ge \log \gamma_n \right ] \right). \IEEEeqnarraynumspace
\end{IEEEeqnarray}
Set 
\begin{IEEEeqnarray}{rCl}\label{eq:gn}
\log \gamma_n = n I(X;Y) + \sqrt{n\mathsf V}Q^{-1}(\alpha_n)
\end{IEEEeqnarray}
with 
\begin{equation}
\alpha_n = \alpha - \frac{0.7975 \mathsf T}{\sqrt{n \mathsf V^3}}  - \frac{\delta}{\sqrt{n}}
\end{equation}
for some $\delta > 0$. 
By employing the strengthened version of the Berry-Esseen theorem \cite{Beeck1972}, we have 
\begin{IEEEeqnarray}{rCl}
\left |\Pr \left [\log \frac{dP_{Y^n|X^n}}{dP_{Y^n}} \ge \log \gamma_n \right ] - \alpha_n \right |\le \frac{0.7975 \mathsf T}{\sqrt{n \mathsf V^3} }.
\end{IEEEeqnarray} 
Consequently 
\begin{IEEEeqnarray}{rCl} \label{eq:65}
\Pr \left [\log \frac{dP_{Y^n|X^n}}{dP_{Y^n}} \ge \log \gamma_n \right ] \le \alpha - \frac{\delta}{\sqrt{n}}.
\end{IEEEeqnarray} 

Substituting \eqref{eq:65} into \eqref{eq:beta}, we have
\begin{IEEEeqnarray}{rCl}
\beta_{\alpha} (P_{Y^n|X^n}, P_{Y^n})  \ge \frac{\delta}{\gamma_n \sqrt{n}},
\end{IEEEeqnarray}
and by \eqref{eq:gn}
\begin{IEEEeqnarray}{rCl} \label{eq:67}
\lefteqn{\log(\beta_{\alpha} (P_{Y^n|X^n}, P_{Y^n}) )}\notag \\
& \ge& \log(\delta) - nI(X:Y) - \sqrt{n \mathsf V}\mathbb Q^{-1}(\alpha_n) - \frac{1}{2} \log (n).
\end{IEEEeqnarray}

By \cite[Theorem 27]{Yuri2012}, every $(\mathsf M, \epsilon)$-code satisfies 
\begin{IEEEeqnarray}{rCl}
\log \mathsf M \le  - \log (\beta_{1-\epsilon} (P_{Y^n|X^n}, P_{Y^n}) ).
\end{IEEEeqnarray}
By \eqref{eq:67} and $\alpha = 1- \epsilon$ and the fact that $\mathbb Q^{-1}(1-x) = - \mathbb Q^{-1}(x)$, 
\begin{IEEEeqnarray}{rCl}
\log \mathsf M
&\le&  nI(X:Y) - \sqrt{nV}\mathbb Q^{-1}(\epsilon + \frac{0.7975 \mathsf T}{\sqrt{n \mathsf V^3}}  + \frac{\delta}{\sqrt{n}}) \notag \\
&& + \frac{1}{2} \log (n) - \log (\delta)
\end{IEEEeqnarray} 
which proves the inequality \eqref{eq:Rl}. Combined with the optimal estimator in \eqref{eq:D}, this proves the theorem.

%-----------------------------------

\end{document}